\documentclass[letterpaper,superscriptaddress,english,aps,prb,twocolumn,floatfix,showpacs, amsfonts, amssymb]{revtex4}
\usepackage{epsfig, graphicx,psfrag,amsmath,amssymb,float}
\usepackage[T1]{fontenc}
\usepackage[latin9]{inputenc}
\usepackage{amsmath}
\usepackage{amssymb}
\usepackage{amscd}
\usepackage{bm}
\usepackage{psfrag}
\usepackage{babel}
\usepackage{graphicx}
\usepackage{pstricks}
\usepackage{sidecap}

\begin{document}
\title{Real-space renormalization group flow in quantum impurity systems: 
\\local moment formation and the Kondo screening cloud}

\author{Andrew K. Mitchell, Michael Becker and Ralf Bulla}

\affiliation{Institute for Theoretical Physics, University of Cologne, 50937 Cologne, Germany}
\date{\today}

\begin{abstract}
The existence of a 
length-scale $\xi_K\sim
1/T_K$ (with $T_K$ the Kondo temperature) has long been predicted in
quantum impurity systems.  
At low temperatures $T\ll T_K$, the standard interpretation is that a 
spin-$\tfrac{1}{2}$ impurity is screened by a surrounding `Kondo cloud' of
spatial extent $\xi_K$. 
We argue that renormalization group (RG) flow between any two fixed points
(FPs) results in a characteristic length-scale, observed in real-space
as a crossover between physical behaviour typical of each FP.  
In the simplest example of the Anderson impurity model, three FPs 
arise; and we show that `free orbital', `local moment' and 
`strong coupling' regions of space can be identified at zero
temperature. These regions are separated by two crossover length-scales
$\xi_{\text{LM}}$ and $\xi_K$, with the latter diverging as the
Kondo effect is destroyed on increasing temperature through $T_K$. One
implication is that moment formation occurs inside the `Kondo cloud',
while the screening process itself occurs on flowing to the strong
coupling FP at distances $\sim \xi_K$. Generic aspects of the
real-space physics are exemplified by the two-channel
Kondo model, where $\xi_K$ now separates `local moment' and 
`overscreening' clouds. 
\end{abstract}

\maketitle

\section{Introduction}
Most fundamental aspects of the Kondo effect are by now
very well understood,\cite{hewson} with various detailed theoretical
predictions having been confirmed directly by experiments on impurity
systems or quantum dot
devices.\cite{hewson,1dot:Goldhaber} 
Key insights
into the underlying physics have been provided by the
renormalization group (RG) concept, where progressive reduction of the
temperature or energy scale results in RG flow between `fixed
points' (FPs) that can be easily identified for a given model. 
In particular, the low-temperature/energy physics 
is governed by RG flow between two fundamental FPs, with
universality arising in terms of the 
crossover energy scale.

The Anderson impurity model (AIM) captures many
generic aspects of quantum impurity physics. \emph{Three} FPs (and
hence two crossover energy scales) arise:  
charge fluctuations on the impurity or quantum dot dominate at the
`free orbital' 
(FO) FP, which describes the high-energy physics. But due to strong
impurity electron correlations, a spin-$\tfrac{1}{2}$ `local
moment' (LM) forms below $T\sim T_{\text{LM}}$ (typically a high-energy
scale). RG flow from this LM FP to the `strong coupling' (SC) FP
occurs on a much lower energy scale $T_K$, and determines the
universal low-energy behaviour. The SC FP itself describes the ground
state in which the incipient moment associated with the LM FP is
dynamically screened by conduction electrons, which 
together form a many-body `Kondo singlet' state.\cite{hewson} 

Surprisingly however, the underlying physics in real-space is still 
somewhat controversial. Various theoretical
studies\cite{PhysRevB.35.8478,PhysRevB.77.180404,PhysRevB.78.195124,PhysRevB.81.045111,springerlink:10.1140/epjb/e2009-00405-y,
sor_aff,PhysRevB.75.041307,
PhysRevB.80.205114,PhysRevB.74.195103,PhysRevLett.97.136604,
PhysRevB.79.100408,2ckaff} have examined real-space
behaviour in a number of impurity models. While they do appear to have
established the existence of a Kondo 
length-scale, $\xi_K \equiv\hbar v_F/k_B T_K$ 
(with $v_F$ the Fermi velocity), the nature of the two spatial regions
separated by it remains unclear. Certain quantities have been studied in
the universal scaling regime,\cite{PhysRevB.35.8478,PhysRevB.77.180404,PhysRevB.78.195124,PhysRevB.81.045111,springerlink:10.1140/epjb/e2009-00405-y,
sor_aff,PhysRevB.75.041307,
PhysRevB.80.205114,PhysRevB.74.195103,PhysRevLett.97.136604,
PhysRevB.79.100408,2ckaff} but the full and exact evolution of real-space
quantities --- and a satisfactory unifying interpretation of the
results --- has not yet been provided.

The prevailing view\cite{AffleckReview10} is that at low temperatures
$T\ll T_K$, a spin-$\tfrac{1}{2}$ impurity forms a singlet by binding to a
surrounding `Kondo cloud' of spatial extent $\xi_K$. 
This has an intuitive appeal from the perspective
of Fermi liquid theory, where one imagines that the impurity is
`invisible' to conduction electrons \emph{outside} the screening
cloud.\cite{AffleckReview10} 
But no clear evidence in support of the 
screening cloud scenario has so far emerged. 
Very recently, refinement of scanning tunneling spectroscopy
techniques reported in Ref.~\onlinecite{natureLDOS} has shown for the
first time that Kondo signatures in the local density of
states (LDOS) appear away from an impurity. Experimental verification
of the elusive Kondo length-scale --- and resolution of the real-space
debate ---  may now finally be within reach.

\begin{SCfigure}[1][b]
  \centering
  \includegraphics[width=42mm]{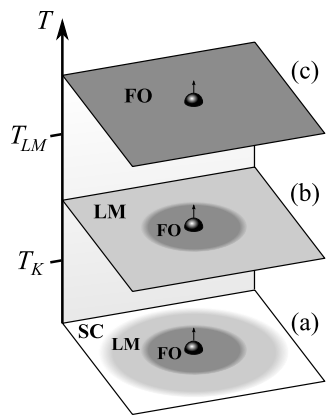}
  \caption{\label{clouds}
Three physical regions arise at $T=0$ in the Anderson impurity model,
corresponding to the FO, LM and SC FPs (a). The crossover
length-scale $\xi_K$ associated with flow to the SC FP diverges as the
Kondo effect is destroyed on increasing temperature through $T_K$ (b),
while $\xi_{\text{LM}}$ diverges for $T\gg T_{\text{LM}}$ (c). The
spin-$\tfrac{1}{2}$ moment forming at $\xi_{\text{LM}}$ 
is screened on flowing to the SC FP at $\xi_K$.}
\end{SCfigure}

Motivated in part by these developments, we consider here
the wider question of how RG flow manifests in real-space.
While necessarily perturbative, a simple intuitive picture is suggested
by well-known scaling arguments applied directly to
real-space,\cite{rsRG}  where notions of RG and universality again
arise. One might anticipate then that this universality could show up in
certain real-space quantities. Indeed, some concrete evidence to
support this general expectation has been provided
theoretically\cite{PhysRevB.35.8478,PhysRevB.77.180404,PhysRevB.78.195124,PhysRevB.81.045111,springerlink:10.1140/epjb/e2009-00405-y,
sor_aff,PhysRevB.75.041307,
PhysRevB.80.205114,PhysRevB.74.195103,PhysRevLett.97.136604,
PhysRevB.79.100408,2ckaff} if not experimentally.

Encouraged by 
these results, we show in this paper that motion away from an
impurity corresponds to RG flow. In fact, we demonstrate explicitly
and exactly that signatures of the \emph{entire} RG flow between all
FPs (whether universal or non-universal) must appear fully in
real-space. Just as flow between a pair of FPs results in a
characteristic energy scale, we show that one can similarly identify
the corresponding \emph{length}-scale, apparent in real-space as a
crossover between physical behaviour typical of each FP. In
particular, this implies that length-scales can be attributed both to
moment formation and moment compensation. In real impurity or quantum
dot systems, the moment screened by the Kondo effect at low
temperatures does not reside on the impurity or dot itself, but is an extended
object which develops in real-space due to electron correlations. 

Further insight is gained by considering the effect on the real-space
physics of increasing temperature. RG flow is cut off at an energy scale
$\sim T$; likewise RG flow is arrested on
the thermal length-scale $\xi_T\propto 1/T$. Consequently, clouds
associated with low-energy FPs reached at larger distances `evaporate'
successively on increasing temperature. Our refined real-space Kondo
screening scenario is summarized schematically in Fig.~\ref{clouds}.

\section{Anderson impurity model in real-space} 
For simplicity and
concreteness, we consider first a semi-infinite tight-binding chain,
$H_{\text{host}}=t\sum_{\sigma}
\sum_{i=0}^{\infty}(c_{i,\sigma}^{\dagger}
c_{i+1,\sigma}^{\phantom{\dagger}} + \text{H.c})$, with a single
Anderson impurity tunnel-coupled to one end:
\begin{equation}
\label{hamand}
H_{\text{And}}=H_{\text{host}} + \epsilon \hat{n}_d + U
\hat{n}_{d,\uparrow}\hat{n}_{d,\downarrow}+ V\sum_{\sigma}(d_{\sigma}^{\dagger}
c_{0,\sigma}^{\phantom{\dagger}} + \text{H.c}),
\end{equation}
where $\hat{n}_d= \sum_{\sigma} \hat{n}_{d,\sigma}\equiv \sum_{\sigma}
d_{\sigma}^{\dagger}d_{\sigma}^{\phantom{\dagger}}$ is the total
number operator for the impurity, and the hybridization strength is
$\Gamma=\pi\rho V^2$ (with $\rho=1/(\pi t)$ the Fermi level density of
states and $2t\equiv 1$ the half-bandwidth). 
As in previous theoretical work,\cite{PhysRevB.77.180404,PhysRevB.78.195124,
PhysRevB.35.8478,PhysRevB.81.045111} we focus here on the `excess' charge density due to the impurity, obtained via
\begin{equation}
\label{dens} 
\Delta n(r,T)=-\tfrac{2}{\pi}\text{Im}\int_{-\infty}^{+\infty}d\omega~f(\omega,T)
~\Delta G_{r,\sigma}(\omega,T),
\end{equation}
with $f(\omega,T)=[1+\text{exp}(\omega/T)]^{-1}$ the
Fermi function and $\Delta G_{r,\sigma}(\omega,T)\equiv
G_{r,\sigma}(\omega,T) - G^0_{r}(\omega)$ the difference in the
site-$r$ Green function with and without the impurity.

\begin{figure}[t]
\begin{center}
\includegraphics*[width=80mm]{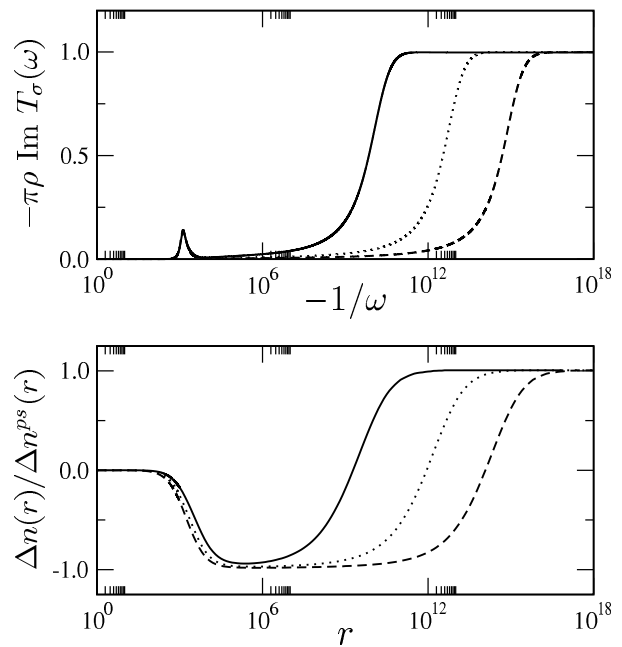}
\caption{\label{rgflow}
Comparison of the spectrum vs inverse frequency (upper panel) and normalized
excess charge density vs distance (lower panel) for the Anderson model
at $T=0$. Plotted for common $10^3V=3$ and $10^3\epsilon=-0.8$, varying
  $10^3U=1$, $1.125$ and $1.25$ (solid, dotted and dashed lines),
  chosen to give exaggerated energy- and length-scale separations for clarity.
}
\end{center}
\end{figure}

The free conduction chain is initially at half-filling, and the
  corresponding no-impurity propagators $G_r^0(\omega)$ are simple
  objects, describing non-interacting electrons. By contrast, the
behaviour of the full Green function $G_{r,\sigma}(\omega,T)\equiv \langle\langle
c_{r,\sigma}^{\phantom{\dagger}} ;c_{r,\sigma}^{\dagger} \rangle
\rangle_{\omega,T}$ is deeply non-trivial due to electron correlations
on the impurity which are ultimately responsible for local moment
formation and thence the Kondo effect. But these correlations
are \emph{local}, and so the subtle real-space behaviour of
$G_{r,\sigma}(\omega,T)$ and $\Delta n(r,T)$ can be
determined from purely local impurity quantities. 
This is most clearly seen using equations of motion,\cite{PhysRevB.81.045111} whence 
for the simple geometry considered here, one readily obtains 
\begin{equation}
\label{eom}
\Delta G_{r,\sigma}(\omega,T) = [\tfrac{1}{t} X(\omega,r)]^2 ~T_{\sigma}(\omega,T).
\end{equation}
All the real-space information about the lattice is contained in
the simple function $X(\omega,r)=[t G_0^0(\omega)]^{r+1}$, where
$G_0^0(\omega)=(\omega-\sqrt{\omega^2-4t^2})/2t^2$ is the free $r=0$
Green function. The spatial dependence of $\Delta
G_{r,\sigma}(\omega,T)$ thus enters only through the power of
$G_0^0(\omega)$, reflecting physically the fact that electrons must
hop $(r+1)$ times to get to the impurity. $T_{\sigma}(\omega,T)=V^2
G_{d,\sigma}(\omega,T)$ is the usual scattering t-matrix, with
$G_{d,\sigma}(\omega,T)\equiv \langle\langle
d_{\sigma}^{\phantom{\dagger}} ;d_{\sigma}^{\dagger} \rangle
\rangle_{\omega,T}$ the impurity Green function. In the trivial
non-interacting limit ($U=0$), the impurity merely gives rise to
potential scattering, and the familiar Friedel density oscillations are
expected.\cite{Friedel58} In this case
$G^{\textit{ps}}_{d,\sigma}(\omega) = [\omega - \epsilon -
V^2G_0^0(\omega)]^{-1}$ independent of temperature. Far from the
impurity, Eqs.~\ref{dens} and \ref{eom} reduce to
$\Delta n^{\textit{ps}}(r,T) =
 -\tfrac{2}{\pi}\text{Im} f(2r,T)\cdot (-1)^r C$, with $f(r,T)= -i\pi
 T/\sinh(\pi r T)$ the Fourier transform of the Fermi function, and 
where $C = 2\epsilon\Gamma/(\epsilon^2+\Gamma^2)
= \sin(2\delta)$ depends purely on the Fermi liquid 
phase shift
$\delta$. At low-temperatures $T\ll 1/r$, one obtains asymptotically
$\Delta n^{\textit{ps}}(r,T) = \tfrac{C}{\pi} (-1)^r/r$, 
while at higher temperatures, density oscillations are exponentially suppressed 
via $\Delta n^{\textit{ps}}(r,T) = 4CT (-1)^r \exp(-2\pi r/\xi_T)$, with
the thermal length-scale $\xi_T =1/T$ so defined, and appearing naturally.

\begin{figure*}[t]
  \centering
\includegraphics*[width=130mm]{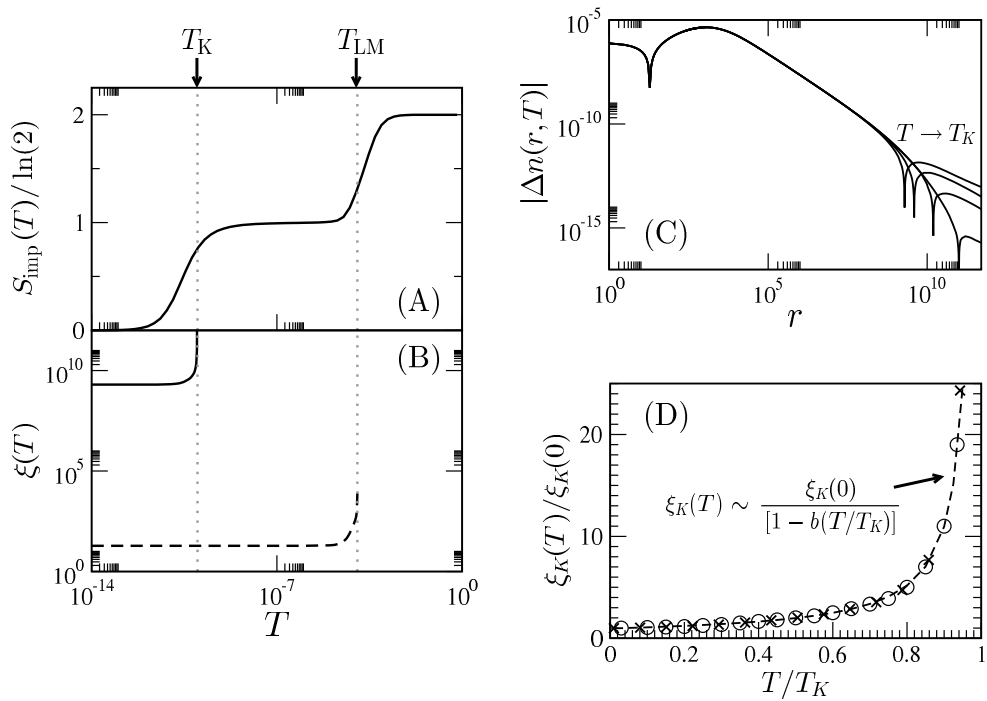}
\caption{\label{tempdep}
Effect of temperature on the Anderson model ($10^{3}V=3$,
$10^{3}\epsilon=-0.8$, $10^{3}U=1$). Panels (A)~and~(B): Comparison of
impurity entropy with length-scales $\xi_K$ (solid lines) and
$\xi_{\text{LM}}$ (dashed) as they evolve with $T$. (C): $|\Delta
n(r,T)|$ vs $r$ for $T/T_K=0.1$, $0.5$, $0.75$ and $0.85$. (D):
$\xi_K(T)/\xi_K(0)$ vs $T/T_K$ for the same system (circles), and
scaling collapse of the divergence to common universal curve for $V=0.05$,
$\epsilon=-0.1$, $U=0.3$ (crosses). Dashed line is a fit to the data.
}
\end{figure*}

The situation of interest is of course the $U>0$ \emph{interacting}
AIM, where $G_{r,\sigma}(\omega,T)$ and hence
$T_{\sigma}(\omega,T)$ now contain information about the Kondo
effect. In this work we employ the numerical renormalization
group (NRG) to obtain accurately $T_{\sigma}(\omega,T)$.\cite{nrg_rev} 
Provided $T\ll T_K$, signatures of Kondo physics and RG flow are
naturally expected
in the Green function or LDOS for any site in
real-space (see Eq.~\ref{eom}), and this has been recently confirmed
experimentally.\cite{natureLDOS} 
But the factorized form of Eq.~\ref{eom} implies that such
measurements cannot reveal a Kondo
\emph{length}-scale, $\xi_K$ (there
can be no universality in terms of $r/\xi_K$).

The key point is that the real-space charge densities are
obtained from a simple integral transformation of the
frequency-resolved t-matrix by Eqs.~\ref{dens} and \ref{eom}.  
Indeed, far from the impurity this transformation reduces to a
Fourier transformation, whose well-defined inverse guarantees the
complete conservation of information for the process
$T_{\sigma}(\omega,T)\leftrightarrow \Delta n(r,T)$. 
In consequence, the full RG structure of
the underlying quantum impurity problem (as manifest in the t-matrix
or LDOS) is \emph{wholly reproduced} in the spatial variation of $\Delta
n(r,T)$. 

This fact is shown at zero-temperature in Fig.~\ref{rgflow}, where the
energy-dependence of the spectrum
$-\pi\rho~\text{Im}~T_{\sigma}(\omega)$ is compared with the
space-dependence of the densities $\Delta n(r) / \Delta n
^{\textit{ps}}(r)$, normalized by the pure
potential scattering contribution at long distances (with
$C=\sin(2\delta)$ and $\delta$ 
now the phase shift of the interacting problem). 
The representative systems plotted in Fig.~\ref{rgflow} have
similar energy scales $T_{\text{LM}}\sim \mathcal{O}(U)$, but widely
differing Kondo scales 
$T_K$. The densities shown in the lower panel exhibit clearly 
RG flow between FPs, now as a function of distance: a non-universal crossover
length-scale $\xi_{\text{LM}}$ characterizes moment formation on
flowing from FO to LM FPs, while $\xi_K$ sets the universal length-scale for
Kondo screening  of this moment on flowing to the SC~FP.

Since the SC FP can be thought of as a free conduction band with one
site removed, a $\pi/2$ phase shift results.\cite{hewson} As shown in
Ref.~\onlinecite{PhysRevB.77.180404} for the Kondo model, this is responsible for a 
sign-change in the Friedel oscillations. Indeed, this provides a
natural explanation of the limiting universal values $\Delta
n(r)/\Delta  n^{\textit{ps}} (r)=-1$ and $+1$ for $r\ll \xi_K$ ($\gg
\xi_{\text{LM}}$) and $r\gg \xi_K$; and the resulting minimum in
$|\Delta n(r)|$ allows direct identification of the Kondo
length-scale.  However, we point out 
that the FO FP can similarly be thought of as a
free conduction band with an additional potential scatterer. The
resulting phase shift $\delta$ also results in a minimum of $|\Delta n(r)|$ at
$r=\xi_{\text{LM}}$. Both length-scales of the Anderson model can be
simply identified in this way, and we find $\xi_K\sim 1/T_K$ and
$\xi_{\text{LM}}\sim 1/T_{\text{LM}}$; consistent with expectations
based on simple scaling grounds.  
In particular, $\xi_K$ grows as the impurity-host coupling
decreases.\cite{hewson} Indeed, Kondo screening
is totally absent in the limit of an uncoupled impurity $V\rightarrow
0$. Here local moment physics persists and $\xi_K$ diverges. 
This implies directly that the spatial region
$r\ll \xi_K$ ($\gg \xi_{\text{LM}}$) is described by the LM FP, and
cannot be as such a `screening cloud'.

One naturally expects the RG structure of the problem to be manifest
also in the temperature-dependence of physical quantities. This is
strikingly apparent in thermodynamics such as the entropy,\cite{nrg_rev} whose impurity contribution flows from
$S_{\text{imp}}(T)=\ln(4)$ at the FO FP to $\ln(2)$ at the LM FP, and
finally $S_{\text{imp}}=0$ at the SC FP, symptomatic of complete Kondo
screening; and as shown in Fig.~\ref{tempdep}(A). 
The corresponding effect of temperature on the real-space physics is
demonstrated in panels (B)--(D). 
Two clear dips in $|\Delta n(r,T)|$ are seen at each temperature in
panel (C), allowing thereby immediate identification of
$\xi_{\text{LM}}(T)$ and $\xi_K(T)$. As the temperature is increased
through $T_K$, we find that $\xi_K(T)$ moves out to \emph{larger}
distances (while $\xi_{\text{LM}}(T)$ remains unaffected). On
destroying the Kondo effect by increasing temperature, one would
expect a `screening cloud' per se to collapse; but understanding $\xi_K$
instead as the length-scale for RG flow from LM to SC FPs, the
divergence of $\xi_K(T)$ indicates simply the spatial persistence of
local moment physics. Temperature cuts off RG flow to the SC FP, and
so there can be no crossover Kondo length-scale to an SC region
of space for $T_K \ll T \ll T_{\text{LM}}$. The LM FP, describing an
unscreened impurity, then pertains for $r\gg
\xi_{\text{LM}}$. 

The characteristic divergence of
$\xi_K(T)$ exhibits universal scaling in terms of $T/T_K$, as shown in
panel (D). This behaviour is found to be described by
$\xi_K(T)\sim \xi_K(0)/[1-b(T/T_K)]$ (dashed line), with
$b=\mathcal{O}(1)$; and as such the diverging length-scale corresponds
roughly to the vanishing energy scale $(T_K-T)$. 
The full evolution of $\xi_K(T)$ and $\xi_{\text{LM}}(T)$ is shown in 
Fig.~\ref{tempdep}(B). The free moment associated with the LM FP is
destroyed by charge fluctuations at the FO FP when $T\gg
T_{\text{LM}}$. $\xi_{\text{LM}}$ thus diverges on 
warming through $T_{\text{LM}}$. The real-space
behaviour in the three distinct temperature regimes is illustrated
pictorially in Fig.~\ref{clouds}.

\section{Multi-channel Kondo model} 
To emphasize the generality of the 
above results, we turn now to 
the $N$-channel Kondo model,\cite{noz,2ckrev} given by
\begin{equation}
\label{mcK}
H_{K}^N = \sum_{i=1}^N \left ( H^i_{\text{host}} + J_i \hat{\textbf{S}}\cdot
\hat{\textbf{s}}_i\right ),
\end{equation}
describing a single spin-$\tfrac{1}{2}$ impurity $\hat{\textbf{S}}$
exchange-coupled to the conduction electron spin-density
$\hat{\textbf{s}}_i$ of channel $i$ at the impurity. 
As before we consider a linear chain geometry, but with potential
scattering\cite{hewson} in each channel now included explicitly since
the impurity is now strictly singly-occupied:   
$H^i_{\text{host}}=\sum_{\sigma} \{
K c_{i,0,\sigma}^{\dagger}c_{i,0,\sigma}^{\phantom{\dagger}} + \sum_{r=0}^{\infty}t(c_{i,r,\sigma}^{\dagger}
c_{i,r+1,\sigma}^{\phantom{\dagger}} + \text{H.c})\}$.  
Straightforward application of the equations of motion yield directly 
an analogue of Eq.~\ref{eom} for each conduction channel $i$:
\begin{equation}
\label{eomK}
\Delta G_{i,r,\sigma}(\omega,T) = \Delta
G^{\textit{ps}}_{r}(\omega) + [\tfrac{1}{t} \tilde{X}(\omega,r)]^2 ~T_{i\sigma}(\omega,T),
\end{equation}
where $\Delta G^{\textit{ps}}_{r}(\omega)$ is the trivial
potential scattering contribution in the $J_i=0$ case,
$\tilde{X}(\omega,r)=X(\omega,r)/[1-K G_0^0(\omega)]$
contains the lattice information, and $T_{i\sigma}(\omega,T)$ are the
t-matrices of the multi-channel Kondo model (again obtained here using
NRG). Friedel oscillations in each channel arise when $K\ne 0$ due to
potential scattering [with $C=\sin(2\delta)=-2Kt/(K^2+t^2)$]. 
But non-trivial many-body effects associated with the
multi-channel Kondo model also appear in the
real-space charge densities $\Delta n_i(r,T)$ for 
$J_i\ne 0$, as now shown for the $N=2$ two-channel Kondo (2CK) 
model.\cite{noz,2ckrev}

\begin{figure}[t]
\begin{center}
\includegraphics*[width=80mm]{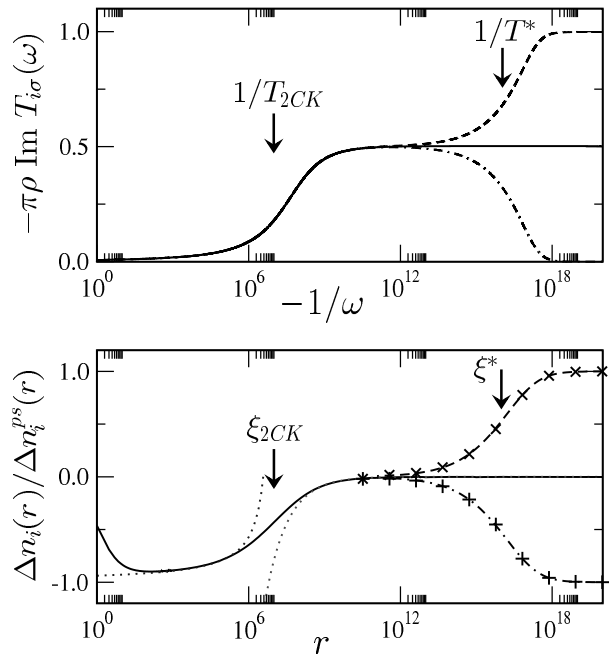}
\caption{\label{specden2ck}
$T=0$ spectrum vs inverse frequency (upper panel) and normalized
density vs distance (lower panel) for the 2CK model with
$(J_1+J_2)=0.2$. Symmetric case $J_1=J_2$ shown as solid lines, while
results for channel 1(2) shown as the dashed(dot-dashed) lines in the
asymmetric case $(J_1-J_2)=10^{-7}$. Common asymptotic behaviour for
$r\ll \xi_{2CK}$ is $\Delta n_i(r)/\Delta n_i^{\textit{ps}}(r) \sim
c/\ln^2(r/\xi_{2CK})-1$ while $\sim (r/\xi_{2CK})^{-1/2}$ behaviour
is observed for $r\gg \xi_{2CK}$ ($\ll \xi^*$). In the asymmetric
case, excellent agreement is seen with the exact result in the vicinity
of the FL crossover, Eq.~\ref{exact_n} (points). 
}
\end{center}
\end{figure}

In the channel-symmetric case $J_1=J_2>0$, RG flow between two FPs
determines the underlying physics. The LM FP
describes the behaviour at 
high energies/temperatures, where the impurity spin is essentially
free. But intrinsic frustration arising from symmetric coupling to two
conduction channels drives the system to the 2CK FP below
$T_{2CK}$, which describes a non-Fermi liquid (NFL) ground state
characterized by `overscreening' of the impurity
spin.\cite{noz,2ckrev} This RG flow is of course clearly manifest in the
behaviour of the t-matrix for each channel, $T_{i\sigma}(\omega)$
(equivalent by symmetry), and so we expect to observe the same flow in
real-space. 

An analog to the single-channel `Kondo screening cloud' has been
suggested for the symmetric 2CK model:\cite{2ckaff} an
`overscreening' cloud of spatial extent $\xi_{2CK}\sim 1/T_{2CK}$
surrounding a partially quenched impurity. But since the ground state
is a NFL, what lies beyond $\xi_{2CK}$? Within the present RG
description, the impurity is surrounded instead by a LM cloud, with
NFL behaviour described by the 2CK FP found in an overscreening cloud
at distances $r\gg \xi_{2CK}$.

While the phase shift concept provides an intuitive explanation for 
physical behaviour in models with Fermi liquid ground
states,\cite{PhysRevB.77.180404} we point out that this is
incidental. No such description in terms of phase shifts exists in the
2CK model, yet the full RG flow and universality of
the problem is again recovered in real-space.

The 2CK FP is in fact the critical point of
Eq.~\ref{mcK}.\cite{noz,2ckrev} For finite channel asymmetry
$(J_1-J_2)\ne 0$, a third FP enters: a Fermi liquid (FL) FP in which 
the impurity spin is completely screened by the more strongly-coupled 
channel. A new energy scale $T^*$ emerges, corresponding to RG flow
from 2CK to FL FPs, and again shows up in the t-matrix for each
channel.\cite{noz,2ckrev,exactT} We find as expected that the
corresponding length-scale $\xi^*\sim 1/T^*$ characterizes real-space
RG flow from a critical 2CK region of space to an FL region. This is
shown explicitly in Fig.~\ref{specden2ck} where we compare (as in
Fig.~\ref{rgflow}) the spectrum vs inverse frequency (upper panel) to
the normalized densities vs distance (lower panel). Two systems are
plotted, with common $K = 0.01$ and $(J_1+J_2) =0.2$, but $(J_1-J_2)
=0$ for the symmetric case (solid lines), while $(J_1-J_2) = 10^{-7}$ for the
asymmetric case (dashed lines for channel 1, dot-dashed lines for
channel 2). The energy scales $T_{2CK}$ and $T^*$ are indicated, as
are the length scales $\xi_{2CK}$ and $\xi^*$.

\begin{figure}[t]
\begin{center}
\includegraphics*[width=85mm]{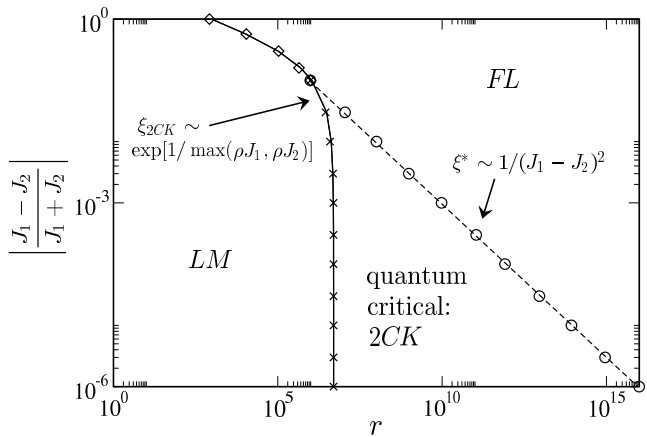}
\caption{\label{2ckpd}
$T=0$ phase diagram for the 2CK model, showing LM, critical 2CK and FL regions
of space. Direct crossover from LM to FL occurs for
large asymmetry (diamond points), but a quantum critical region of
space appears for smaller asymmetry. Two distinct scales are observed
in this regime: $\xi_{2CK}$ (crosses) and $\xi^*$ (circles). Solid
line is fit to expected behaviour $\xi_{2CK}\equiv 1/T_{2CK}$ while dashed line
is $\xi^*\equiv 1/T^*$.
}
\end{center}
\end{figure}

Characteristic asymptotic behaviour in the vicinity of $\xi_{2CK}$ (given in
the caption to Fig.~\ref{specden2ck}, and shown as the dotted lines) can
be extracted from Eqs.~\ref{eomK},~\ref{dens} using the known
asymptotics of the t-matrix for the symmetric 2CK model. Indeed, the
full density crossover from 2CK to FL FPs for $r\gg \xi_{2CK}$
can be calculated using the exact t-matrix announced recently in
Ref.~\onlinecite{exactT}. Our exact result at zero-temperature and small $K$
follows as
\begin{equation}
\label{exact_n}
\frac{\Delta n_i(r) }{\Delta n_i^{\textit{ps}}(r)} ~ \overset{r\gg
  \xi_{2CK}}{\sim} ~ \pm \sqrt{\tfrac{2}{\pi}}
\exp(r/\xi^*)\sqrt{r/\xi^*} \text{K}_0(r/\xi^*),
\end{equation}
where $\text{K}_0$ is the modified Bessel function of the second kind,
and the $+$($-$) sign is used for the density in channel
1(2). The essentially perfect agreement between Eq.~\ref{exact_n}
(points) and the full numerical solution (lines) is shown in the lower
panel of Fig.~\ref{specden2ck}. Interestingly, this result indicates
that the density crossover approaching the FL FP of the 2CK model is
closely related to the spatial crossover in the magnetization resulting 
from a boundary magnetic field acting in the Ising model.\cite{exactT} 

Finally, we plot the phase diagram for the 2CK model in
Fig.~\ref{2ckpd}. Three regions of space can be identified at zero
temperature, corresponding to LM, critical 2CK and FL FPs. The
crossover length-scales $\xi_{2CK}\sim 1/(J_1-J_2)^2$ (dashed line) and
$\xi_{2CK}\sim \exp[1/\max(\rho J_1, \rho J_2)]$ (solid line) are
anticipated from the well-known variation of the energy scales
$T_{2CK}$ and $T^*$ in the asymmetric 2CK model,\cite{noz,2ckrev} and
agree excellently with explicit calculation (points). Even in the
`standard' case where some degree of channel asymmetry is present in
the model (in which case the ground state is a Fermi liquid), the
non-Fermi liquid correlations at finite energies give rise to an
intermediate region in space describing quantum critical behaviour.

\section{Conclusion} 
Quantum impurity problems are generically
described in terms of an RG framework: as a function of energy or
distance. Specifically, static charge density oscillations away from
an impurity in real-space are simply related by integral
transformation to the energy-resolved scattering t-matrix --- a
dynamic quantity, itself related to the impurity LDOS. As such, the
well-known and rich behaviour associated with RG flow between fixed
points is wholly reproduced in the densities. Indeed, this is not
confined to the simplest 1-dimensional geometry considered explicitly
here. Although real-space behaviour does of course depend on the
particular physical system under consideration, the underlying RG
structure of the problem must still appear. 

One implication is that
the `Kondo cloud' surrounding an impurity, typically defined as the
region $r< \xi_K$, actually corresponds mainly to the LM FP. The screening
process itself occurs on flowing to the SC FP at $r\sim \xi_K$. 

\begin{acknowledgments}
We acknowledge funding from the Deutsche Forschungsgemeinschaft
through SFB 608 and FOR 960. 
 \end{acknowledgments}

\end{document}